\documentclass[10pt]{article}
\usepackage{pslatex} 
\usepackage[paperheight=22.86cm,paperwidth=15.24cm,left=1.27cm,right=1.27cm,top=1.6cm,bottom=1.6cm,headheight=1in]{geometry}
\usepackage[utf8]{inputenc}
\usepackage[T1]{fontenc}
\usepackage[hidelinks]{hyperref}
\usepackage{graphicx} 
\usepackage{enumerate} 
\usepackage[font={small,sf},format=plain,labelfont=bf,up]{caption}
\usepackage{sectsty}
\sectionfont{\fontsize{12pt}{12pt}\selectfont} 
\usepackage{titlesec} 
\titlelabel{\thetitle.\quad}

\usepackage{caption}
\usepackage{subcaption}
\usepackage{enumitem}
\usepackage{hyperref}
\usepackage{multirow}
\usepackage{url}
\usepackage{float}
\usepackage[numbers]{natbib}
\begin{document}

\begin{center}
    {\fontsize{14pt}{14pt}\selectfont \textbf{Bayesian Evaluation of User App Choices in the Presence of Risk Communication on Android Devices}\par}
    \vspace{\baselineskip}

    $~B. Momenzadeh^1,~S. Gopavaram^1, ~S. Das^{1,2},~and~L. J Camp^1$
    %\vspace{\baselineskip}
    
    {\fontsize{10pt}{10pt}\selectfont 1. Indiana University Bloomington\par}
    {\fontsize{10pt}{10pt}\selectfont 2. University of Denver\par}
    {\fontsize{10pt}{10pt}\selectfont (smomenza, sgopavar, sancdas, ljcamp)@iu.edu\par}
    
\end{center}
\linespread{0.7}
\section*{Abstract}

{\fontsize{9pt}{9pt}\selectfont 
In the age of ubiquitous technologies, security- and privacy-focused choices have turned out to be a significant concern for individuals and organizations. Risks of such pervasive technologies are extensive and often misaligned with user risk perception, thus failing to help users in taking privacy-aware decisions. Researchers usually try to find solutions for coherently extending trust into our often inscrutable electronic networked environment. To enable security- and privacy-focused decision-making, we mainly focused on the realm of the mobile marketplace, examining how risk indicators can help people choose more secure and privacy-preserving apps. We performed a naturalistic experiment with $N=60$ participants, where we asked them to select applications on Android tablets with accurate real-time marketplace data. We found that, in aggregate, app selections changed to be more risk-averse in the presence of user risk-perception-aligned visual indicators. Our study design and research propose practical and usable interactions that enable more informed, risk-aware comparisons for individuals during app selections. We include an explicit argument for the role of human decision-making during app selection, beyond the current trend of using machine learning to automate privacy preferences after selection during run-time.\par}

\section*{Keywords}
{\fontsize{10pt}{10pt}\selectfont Mobile App Permissions, Android, Risk Communication, Human-Centered Privacy and Security, Mobile Security.\par}
\vspace{-4mm}
\section{Introduction}
Permissions models are an excellent initiative to inform smartphone users of the services that each application might access. However, research has shown that they have failed to consistently communicate the privacy and security risks of apps on mobile platforms~\cite{agarwal2013protectmyprivacy,felt2012android,kelley2012conundrum}. Currently, many researchers are discarding permissions as futile user communication, focusing on implicit instead of explicit choices and using machine learning or agent-based permissions management after installation~\cite{olejnik2017smarper,wijesekera2017feasibility}. Not only does much research in this area focus on building machine learning tools that regulate resources accessed by apps during runtime, but Android OS has also shifted from app permission manifests to runtime permissions. Mitigating privacy risks for apps during runtime is essential, and much of this mitigation must be automated. However, an automated system during runtime has its own limitations. 
 
Our work motivation is to determine if a multi-level communication system can support explicit individual decision-making during app selection. In addition to supporting individual autonomy, privacy-aware decision making at the time of application election offers promise for the entire ecosystem. Supporting individual risk-aware decisions in app selection could enable app providers to differentiate themselves in the app marketplace and provide developers with an incentive to consider user privacy when building apps. In this paper, we focus on enhancing the decision-time communication of risks to the user. We built a risk-indicative warning system and tested it with an operational app store in a natural environment. This warning system is built upon the findings of previous research in usable security on mobile devices and behavioral psychology.

The essential contribution of our work is an empirical illustration of the changes in participants' decision-making when provided with simple, timely, comprehensible warnings. Instead of removing permissions interactions, an alternative approach is to improve the communication and design aspect to enable users to take privacy-aware and security-aware decisions. Specifically, we illustrate the efficacy of a multi-level system where information is immediately available and summarized, with the option of searching for additional information. As this is a recommendation for design in general practice and for warning systems specifically, this is not surprising~\cite{wogalter1999warnings}. In addition we use a Bayesian experiment design and analysis to compare the distribution of app selections from participants in our market to those in the standard app market. The purpose of using a Bayesian approach is to test the interaction in a noisy, confounded naturalistic environment and to provide a stronger confidence measure than a traditional means comparison.
\vspace{-4mm}
\section{Background Motivation}
\subsection{Permission Models on Mobile Phones}
To develop our warning system, we leveraged the Android permissions model that was used before Android OS moved to the runtime model (which was previously only used by iOS). Instead of presenting the list of permissions immediately, we added a layer of interaction that summarizes the risk of the agreed-upon permissions by the users. 

Empirical research has found a significant lack of understanding, not only about the implications of providing sensitive permissions but also about the underlying meaning of permissions~\cite{felt2012android,kelley2012conundrum}. Smartphone users are mostly unaware of the resources accessed by apps~\cite{mylonas2013delegate}. For permission manifests used in Android, repeated research has shown that people usually ignore or pay little attention to them; for example, a series of online surveys and laboratory studies conducted by Felt et al. found that only 17\% of the participants paid attention to permissions during app installation~\cite{felt2012android}. Another study conducted by Rajivan et al. four years later found that only 13\% of the participants viewed the permissions by clicking on them~\cite{rajivan2016influence}.

In recognition of the fact that previous permissions models were inadequate, there has been a move to automate permissions decisions based on machine learning models of observed user behavior. Models of user preferences may be driven by background observations, possibly augmented by explicit queries about acceptable data use~\cite{olejnik2017smarper,wijesekera2017feasibility}. The addition of machine learning mitigates risk, but it does not enable purposeful choice. Those who value their privacy are unable to make privacy-preserving app selections, as there is a lack of adequate decision-making support at the moment of selection~\cite{bohme2007viability}. 
\vspace{-4mm}
\subsection{Visual Indicators}
We based the design of our visual warning risk-indicator for aggregate privacy on previous work and chose padlocks. We decided to frame the indicator positively, so more padlocks implied a lower risk. For that reason, we refer to these ratings as \textit{privacy ratings}. We considered the five principles proposed by Rajivan et al.~\cite{rajivan2016influence}. First we selected \textit{icons aligned with user mental models of security}, meaning we selected the widely-used lock icon from HTTPS. Second, given that \textit{privacy communicating icons should be in terms of privacy offered by the app/software}, we based ratings on the permissions. Third, we made the \textit{scale of privacy communicating icons consistent with other indicators}. Fourth, and this inherently aligns with our design, \textit{icons should be presented early in the decision-making process, while people compare apps to choose and install}. And the fifth principle, that \textit{privacy communication should be trustworthy}, was embedded in our use of permissions for rating the apps.
\vspace{-4mm}
\section{Methodology}
The goal of the experiment is to investigate if the introduction of proposed visual indicators in an actual PlayStore would change user app selections. Thus, we built an alternate PlayStore and asked participants to select multiple apps from different categories. We then ranked the apps presented to the users based on the number of downloads they received in the experiment. We compared these rankings against the download-based rankings in the actual PlayStore. Through this study design, we aimed at answering the following research questions (RQ):
\begin{description}
    \item[RQ1] In the absence of differing privacy options, do our participants make choices that are indistinguishable from the Android Marketplace?
    \item[RQ2] When the functionality of the apps is the same, but the privacy options differ, do our participants make choices that are indistinguishable from the Android Marketplace?
    \item[RQ3] When both functionality and privacy options vary, do our participants make choices that are indistinguishable from the Android Marketplace?
\end{description}
\vspace{-4mm}
\subsection{Alternate PlayStore}
\label{sec:AlternativePlayStore}
\begin{figure}[h]
    \centering
\begin{minipage}{0.45\textwidth}
    \includegraphics[scale=0.3]{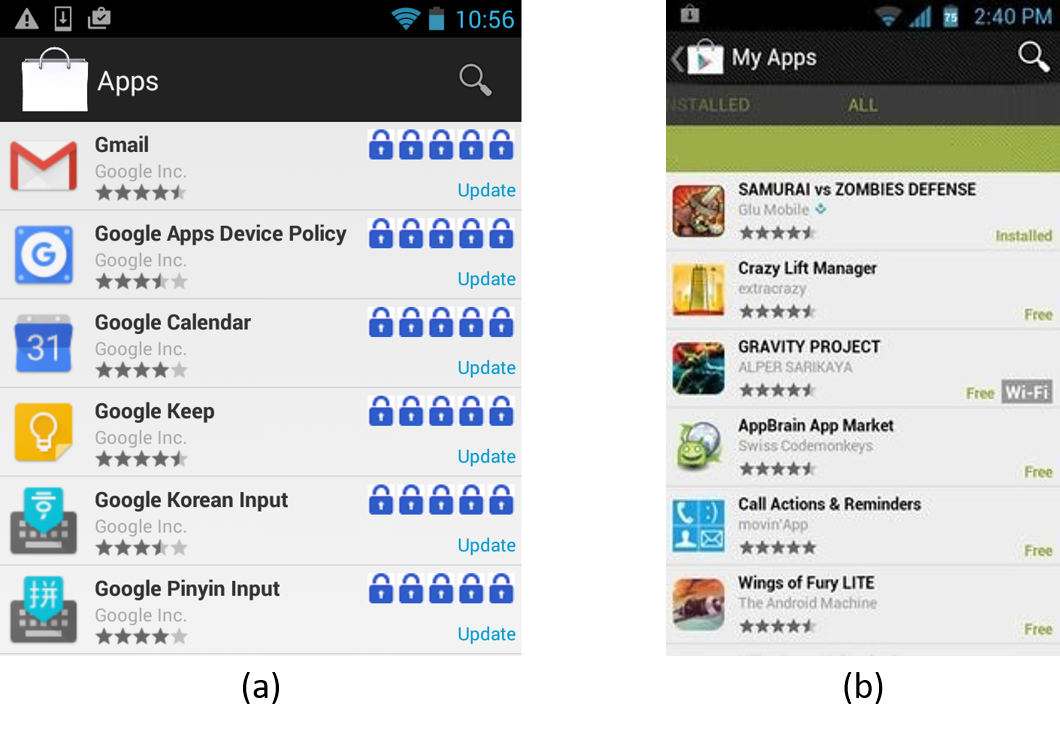}
    \caption{(a) \textit{List of apps} page on the alternate PlayStore with privacy ratings on the alternate PlayStore. (b) \textit{List of apps} page with no risk score on the actual PlayStore}
    \label{fig:ListOfApps-APS}
\end{minipage}
\hspace{1cm}
\begin{minipage}{0.45\textwidth}
    \includegraphics[scale=0.24]{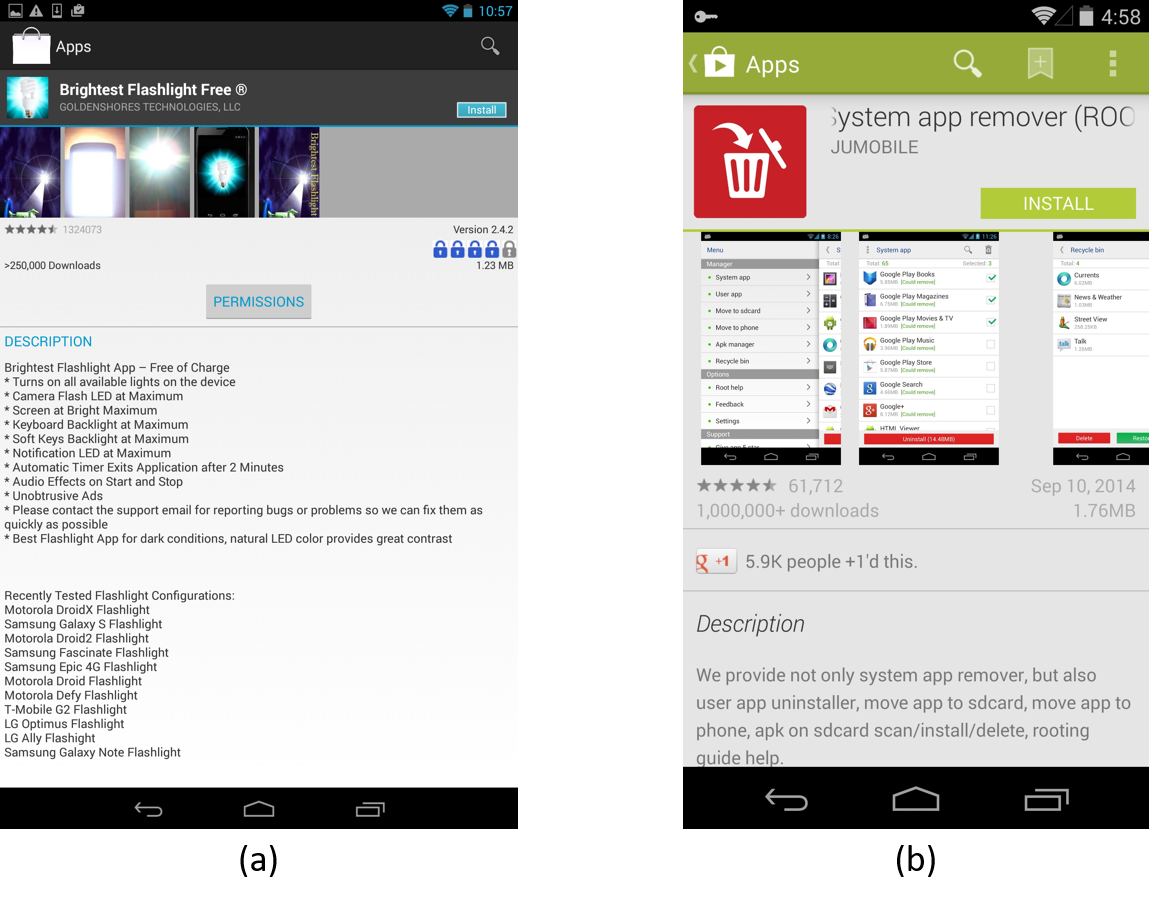}
    \caption{(a) \textit{App description} page on the alternate PlayStore with privacy ratings on the alternate PlayStore. (b) \textit{App description} page with no risk score on the actual PlayStore}
    \label{fig:AppDescription-APS}
\end{minipage}
\vspace{-5mm}
\end{figure}

To answer the research questions mentioned above, we built a functional app store with real-world applications, app ratings, and download counts. The user interface for our app store resembled that of Google's PlayStore on Android Jelly Bean (Version 4.1). Unlike Google's PlayStore, our app store presented users with visual indicators for aggregate risk (privacy ratings). We derived the privacy ratings from PrivacyGrade~\cite{lin2012expectation,lin2014modeling}~\footnote{\url{http://privacygrade.org/}}. PrivacyGrade generated privacy grades ranging from A through D for apps on Android. We retrieved the privacy grade and converted it into a numerical rating between 1 and 5. Since we use positive framing, a privacy rating of 5 is equivalent to an A grade. The privacy rating is presented on both the \textit{list of apps} page and the \textit{app description} page. Figures~\ref{fig:ListOfApps-APS} and~\ref{fig:AppDescription-APS} show the \textit{list of apps} and the \textit{app description} pages respectively alongside their counterparts from the actual PlayStore. We added a button at the top of the description page which would show permissions to the participants. We used this button to track which participants viewed the permissions.

We built the alternative PlayStore by modifying the code of \textit{BlankScore~\footnote{\url{https://github.com/mar-v-in/BlankStore}}} (An open-source Google PlayStore client) and used an open-source API to query Google's servers for information. The alternative PlayStore enabled us to provide accurate user ratings, download counts, descriptions, and a list of permissions of apps. Additionally, search results for applications on the alternative PlayStore were the same as the results on the actual PlayStore, including the order of presentation of apps.
\vspace{-4mm}
\subsection{Study Design}
We recruited a total of $N=60$ participants for the experiment through our outreach at the public library and the local farmers market to obtain socio-economically and culturally homogeneous population. A core design goal was to make the experimental interaction as close as possible to the experience of interacting with the Android PlayStore. We installed the alternate PlayStore on Nexus 7 tablets and provided those tablets to our participants. We then provided each of our participants with a list of keywords to search for on the alternate PlayStore. These keywords correspond to the app categories we chose for our experiment. Each search provided a list of up to 16 apps for a given category, and we asked the participants to select and download 4 of them. To make sure all the participants saw the same results, we ensured that each participant used the same category names. The search results for all the keywords on the alternate PlayStore were identical to the ones generated by the actual PlayStore. We did not describe the purpose of the experiment, mention security, nor describe the indicators to the participants beforehand.
\vspace{-4mm}
\subsection{Statistical Analysis Approach}
We used a clinical research model with an observational study by selecting a subset of participants and exposing them to an experimental condition then comparing their outcomes with the large known set of results without the condition~\cite{rohrig2009types}. We compared the means of the two groups, using a posthoc Tukey pairwise comparison. We include these results for each category for the ease of comparison with other work; however, we also argue that the lack of nuance in means comparisons argues for the use of a Bayesian approach. The Kruskal-Wallis Test shows the significance of differences in weighted means of privacy ratings for the four categories, which are: 0.005 for Games; 0.53 for Flashlights; 0.02 for Photos; and 0.28 for Weather. The results of this comparison show the significance of the differences between the mean privacy rating of apps chosen by those using our experimental PlayStore and the mean privacy rating of apps chosen through the actual Android PlayStore. We calculated a region of practical equivalence (ROPE) based on a Highest Density Interval (HDI) of 95\%, which means the area that contains the 95\% most credible values for participants' choices have the same distribution of the users' choices in the PlayStore. The comparison is between the behavior of our experimental sample and the behavior of people using the regular Android PlayStore. The advantages of an analysis using a Bayesian approach are that it integrates historical information and that it is valid with a small sample size. Other advantages are that the Bayesian analysis requires no assumptions about normality or distribution of the data. When examining the difference of means graphs we provide in reporting our Bayesian analysis, the dotted line on zero marks the point where the distributions match. The black line underneath the bars defines the 95\% HDI area. The numbers on either side of the black line to specify the start and end threshold of the 95\% interval. The bars show the distribution of the Difference of Means data. We will talk about each graph individually below as we report the results for each of the four categories.
\vspace{-4mm}
\subsection{Results}
\label{sec:Exp1-Results}
Out of the $60$ participants, 58\% were male, and 42\% were female. After completing the experiment, all the participants in the study were asked to answer questions related to their app installation behavior. One of the questions asked them about the criteria they considered when selecting an application. In response to this question, 48\% of the participants stated that they prioritized an app's features over other criteria when selecting an app to install. After that, the popularity ranked second, and friends' suggestion was the third choice. The other criteria, in this case, were ads, permissions, rank, reviews, and design. The survey inquired about permissions behaviors, asking participants about how often they checked an app's permissions. To this, 22\% of the participants stated that they check permissions ``almost every time\rq\rq~or ``always\rq\rq~when installing an app. We also asked participants if they had previously refused to continue with the installation of an app because of its permissions. 79.6\% of the participants stated that they had refused to install an app because of the permissions it requested in their real life. However, in practice, there was only one instance where a participant did not continue with the installation of an app because of its permissions (or after viewing the permissions). Only 7\% of the total installations preceded with a check of the app's permissions in our experiment. This discrepancy between the observed and stated behavior is consistent with previous research studies~\cite{rajivan2016influence,yang2012short}.

We also investigated if the addition of aggregate risk information was cognitively burdensome for our participants. Therefore, we used the NASA TLX instrument to measure the mental workload involved in using the alternate PlayStore to select apps~\cite{hart1988development}. The results indicate that the majority of the participants (78\%) found the workload to be minimal.
\vspace{-3mm}
\subsubsection{Research Question 1: Are Participants Representative?}
We chose the Flashlight category to address this research question. It is not that flashlights are particularly safe and secure; rather, such apps all have the same level of privacy in terms of permissions. Table~\ref{table:flashlights} shows both participant selections, PlayStore selections, and the similar privacy ratings of all of the apps. All flashlight apps also have the same functionality. When there is no difference in the privacy indicators nor the functionality of the apps, we wanted to explore whether the selections of our participants were different from those in the PlayStore. Using Bayesian analysis we show that our participants were indistinguishable from a random sample of people selecting Android apps.
\begin{table}
\centering
\resizebox{0.7\textwidth}{!}{%
\begin{tabular}{|l|c|c|c|c|}
\hline
App Name & \multicolumn{1}{l|}{Downloads} & \multicolumn{1}{l|}{Locks} & \multicolumn{1}{l|}{Exp. Rank} & \multicolumn{1}{l|}{PlayStore Rank} \\ \hline
Super-Bright LED Flashlight & 38 & 5 & 1 & 1 \\ \hline
Color Flashlight & 34 & 5 & 2 & 3 \\ \hline
Tiny Flashlight + LED & 26 & 5 & 3 & 2 \\ \hline
Brightest Flashlight Free & 20 & 4 & 4 & 4 \\ \hline
Flashlight Galaxy S7 & 16 & 5 & 5 & 10 \\ \hline
Flashlight Galaxy & 16 & 5 & 5 & 9 \\ \hline
Brightest LED Flashlight & 15 & 5 & 7 & 5 \\ \hline
Flashlight & 12 & 5 & 8 & 11 \\ \hline
High-powered Flashlight & 11 & 5 & 9 & 6 \\ \hline
Flashlight Widget & 7 & 5 & 10 & 12 \\ \hline
FlashLight & 6 & 5 & 11 & 7 \\ \hline
Flashlight for HTC & 5 & 5 & 12 & 13 \\ \hline
Flashlight & 3 & 5 & 13 & 8 \\ \hline
\end{tabular}%
}
\caption{Flashlight Category by order of Downloads in the Experiment: Apps' Rank in the PlayStore, Downloads in the Experiment, and Privacy Rating (Locks)}
\vspace{-2mm}
\label{table:flashlights}
\end{table}

\begin{table}
\centering
\begin{tabular}{| c |}
\hline
\includegraphics[width=0.4\textwidth]{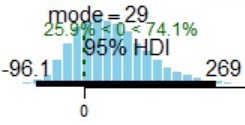} \\
Flashlight Category \\
\hline
\end{tabular}
\captionof{figure}{Regions of Practical Equivalence for the selection of flashlight apps (RQ1). The comparison of the selections in our sample ($\mu_{1}$) and the selections of our participants ($\mu_{2}$).  The graph shows the ROPE for the difference of means ($\mu_{1}-\mu_{2}$), showing that our participants' selections were indistinguishable from a random sample of Android app purchasers. }
\label{figure:Flashlight}
\vspace{-5mm}
\end{table}

To address the frequentist results first, the weighted average privacy ratings of the flashlight apps in the PlayStore is 4.94. The weighted app ratings were 4.52 and 4.51 for the store and the experimental participants, respectively. In Figure~\ref{figure:Flashlight}, 0 is marked with a dotted line. Zero falls near the center of the region of practical equivalence. The entire HDI is within the ROPE, so the difference is practically equivalent to the null value. In the case of the Flashlight apps, our participants were statistically indistinguishable from a random sample of the selections made in the larger PlayStore. This verifies that, in the absence of differing privacy ratings, our participants' choices were indistinguishable from those of a random sample of Android users.
\vspace{-3mm}
\subsubsection{Research Question 2: Similar Functionality and Different Privacy Rating}
To answer RQ2, we used the apps in the Weather category. In many cases, we expected that the individuals would trade privacy or security for some other feature-based benefit. Given the difficulty in measuring how individuals value risk avoidance, we sought a category with little functional variance and high variability in information risk. To begin with an illustrative frequentist comparison, the weighted average privacy ratings of the weather apps in the PlayStore were 4.26 and 4.25 for our participants. The weighted average app ratings were 4.39 for both PlayStore users and experimental participants. The Kruskal-Wallis difference in means had a p-value of 0.28. (Note that Kruskal-Wallis examines the contrast of the ways, while a Bayesian approach considers the likelihood of a distribution).
\begin{table}[htpb]
\centering
\resizebox{0.7\textwidth}{!}{%
\begin{tabular}{|l|c|c|c|c|}
\hline
App Name & \multicolumn{1}{l|}{Downloads} & \multicolumn{1}{l|}{Locks} & \multicolumn{1}{l|}{Exp. Rank} & \multicolumn{1}{l|}{PlayStore Rank} \\ \hline
\small{Weather - The Weather Channel} & 40 & 4 & 1 & 1 \\ \hline
AccuWeather & 31 & 5 & 2 & 2 \\ \hline
Yahoo Weather & 27 & 5 & 3 & 5 \\ \hline
MyRadar Weather Radar & 27 & 5 & 3 & 10 \\ \hline
Weather Underground & 19 & 5 & 5 & 11 \\ \hline
Weather by WeatherBug & 16 & 3 & 6 & 6 \\ \hline
Weather \& Clock Widget Android & 14 & 4 & 7 & 4 \\ \hline
Transparent clock \& weather & 11 & 3 & 8 & 6 \\ \hline
NOAA Weather Unofficial & 7 & 4 & 9 & 12 \\ \hline
Go Weather Forecast & 5 & 4 & 10 & 3 \\ \hline
Weather Project & 5 & 1 & 10 & 15 \\ \hline
Weather, Widget Forecast Radar & 3 & 4 & 12 & 8 \\ \hline
Weather Project & 2 & 1 & 13 & 14 \\ \hline
iWeather-The Weather Today & 2 & 1 & 13 & 13 \\ \hline
Weather & 1 & 4 & 15 & 9 \\ \hline
\end{tabular}%
}
\caption{Weather Category by order of Downloads in the Experiment: Apps' Rank in the PlayStore, Downloads in the Experiment, and Privacy Rating (Locks)}
\label{table:Weather}
\end{table}
\begin{table}
\centering
\begin{tabular}{| c |}
\hline
\includegraphics[width=0.6\textwidth]{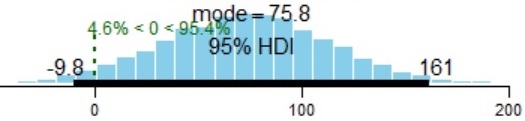} \\
Weather Category \\
\hline
\end{tabular}
\captionof{figure}{Regions of Practical Equivalence, showing the difference between our participants ($\mu_{1}$)  and the selections in the Android marketplace ($\mu_{2}$). The graph shows the ROPE for the difference of means ($\mu_{1}-\mu_{2}$), showing less than 5\% overlap with 0 region and an unskewed distribution. }
\label{fig:weather}
\vspace{-5mm}
\end{table}
The dominance of the most popular weather app, with a privacy rating of four, results in a slight skewing of the results. In many contexts, it is well-understood that people select the first choice on a list or go with defaults~\cite{johnson2002defaults,lai2006internet}. The overall difference in the means between weather apps is shown in Figure~\ref{fig:weather}. This shows little overlap between the distribution of selected weather apps between our participants and the distribution of participants in the PlayStore. That is, the likelihood that the selection of apps by our participants is an unbiased distribution resulting from a sample of the prior distribution, as shown by the PlayStore, is practically equivalent to the null value. This result indicates that the distribution is biased, and thus our warning visual risk-indicator affected our participants' choices.
\vspace{-7mm}
\subsubsection{Research Question 3: Varying Functionality and Privacy Rating}
Our other two categories, Photos and Games, were used to answer this research question. In photos, there is more variance in functionality than in weather or flashlight apps. Photo apps coordinate with different services (e.g., Instagram or Facebook), offer different filters (e.g., glitter, party hats, sepia tones), different functionality (e.g., annotating), and different sharing modes. The weighted average privacy ratings of the game apps in the PlayStore and of the choices of participants were both 4.93. The weighted app ratings were 4.43 and 4.34 for the store and the experimental participants, respectively. The Bayesian analysis of the Games category is shown in Figure~\ref{fig:games}. The dotted line falls into the 95\% HDI. The volume of the ROPE that intersects with the likelihood of this being the practical equivalent of a random, unbiased sample of the prior known PlayStore distribution is 4.6\%, approaching 5\%. The initial value of the HDI (-58.5) is comparable to the distance between the zero and mode, which falls on 102. The average privacy rating of the photos apps in the PlayStore is 4.69. The weighted average privacy ratings of the choices of participants were 4.97. The weighted app ratings were 4.39 and 4.41.

\begin{table}[htpb]
\centering
\resizebox{0.7\textwidth}{!}{%
\begin{tabular}{|l|c|c|c|c|}
\hline
App Name & \multicolumn{1}{l|}{Downloads} & \multicolumn{1}{l|}{Locks} & \multicolumn{1}{l|}{Exp. Rank} & \multicolumn{1}{l|}{PlayStore Rank} \\ \hline
\small{Fruit Ninja Free} & 39 & 5 & 1 & 2 \\ \hline
Subway Surfers & 23 & 5 & 2 & 1 \\ \hline
Super Smash Jungle World & 22 & 5 & 3 & 8 \\ \hline
PAC-MAN & 20 & 5 & 4 & 5 \\ \hline
Wheel of Fortune Free Play & 16 & 5 & 5 & 13 \\ \hline
Color Switch & 15 & 5 & 6 & 7 \\ \hline
Piano Tiles 2™ & 15 & 5 & 6 & 4 \\ \hline
slither.io & 12 & 5 & 8 & 3 \\ \hline
Rolling Sky & 11 & 5 & 9 & 6 \\ \hline
Block! Hexa Puzzle & 4 & 1 & 10 & 9 \\ \hline
Flip Diving & 3 & 5 & 11 & 10 \\ \hline
Battleships - Fleet Battle & 2 & 5 & 12 & 17 \\ \hline
Snakes \& Ladders King & 2 & 5 & 12 & 11 \\ \hline
Board Games & 1 & 5 & 14 & 13 \\ \hline
Best Board Games & 1 & 5 & 14 & 15 \\ \hline
Checkers & 1 & 5 & 14 & 12 \\ \hline
Mancala & 1 & 3 & 14 & 16 \\ \hline
\end{tabular}%
}
\caption{Games Category by order of Downloads in the Experiment: Apps' Rank in the PlayStore, Downloads in the Experiment, and Privacy Rating (Locks)}
\label{table:Games}
\end{table}
\begin{table}
\centering
\begin{tabular}{| c |}
\hline
\includegraphics[width=0.5\textwidth]{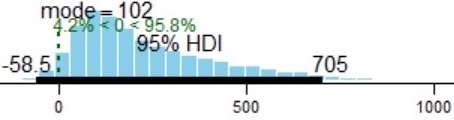}\\
Games Category \\
\hline
\end{tabular}
\captionof{figure}{ROPE for the difference of means ($\mu_{1}-\mu_{2}$) for our participants and the Android marketplace in selection of game apps. Note the distribution of probabilities is highly skewed, decreasing confidence, while the overlap is 4.2\% (RQ3)}
\label{fig:games}
\vspace{-5mm}
\end{table}

In the case of photo apps, the distribution of app ratings and risk was such that individuals could mitigate risk without sacrificing any benefits. With photo applications, participants chose more secure apps over other more popular apps with more downloads, more familiarity, and more popular design. \textit{PicsArt Photo Studio and Collage} was particularly selected by only three of our participants in our experiment for the photos category, while it was the second-ranked app in terms of the number of downloads with this search term in Google PlayStore. The results for the Photos category is quite similar to the Weather category. Zero falls at the beginning of the HDI interval. However, the distance to the start of the range is insignificant, and the distance to mode is also significant. This implies that we have influenced participants' decisions in this category as well and that participants' choices could be distinguished from options in the PlayStore. The results show that the likelihood that the parameters that characterize the distribution of the choices by our participants cannot reasonably be considered the same as the parameters that characterize the distribution of apps chosen by those in the PlayStore.

\begin{table}[htpb]
\centering
\resizebox{0.7\textwidth}{!}{%
\begin{tabular}{|l|c|c|c|c|}
\hline
App Name & \multicolumn{1}{l|}{Downloads} & \multicolumn{1}{l|}{Locks} & \multicolumn{1}{l|}{Exp. Rank} & \multicolumn{1}{l|}{PlayStore Rank} \\ \hline
Google Photos & 39 & 5 & 1 & 1 \\ \hline
PhotoDirector Photo Editor App & 25 & 5 & 2 & 9 \\ \hline
Photo Lab Picture Editor FX & 24 & 5 & 3 & 5 \\ \hline
Gallery & 23 & 5 & 4 & 10 \\ \hline
Photo Editor Pro & 20 & 5 & 5 & 4 \\ \hline
A+ Gallery Photos & 19 & 5 & 6 & 12 \\ \hline
Photo Collage Editor & 17 & 5 & 7 & 5 \\ \hline
PhotoGrid \& Photo Collage & 15 & 5 & 8 & 3 \\ \hline
Toolwiz Photos - Pro Editor & 13 & 5 & 9 & 11 \\ \hline
Photo Editor Collage Maker Pro & 9 & 5 & 10 & 7 \\ \hline
PicsArt Photo Studio & 3 & 3 & 11 & 2 \\ \hline
Phonto - Text on Photos & 1 & 5 & 12 & 8 \\ \hline
\end{tabular}%
}
\caption{Photos Category by order of Downloads in the Experiment: Apps' Rank in the PlayStore, Downloads in the Experiment, and Privacy Rating (Locks)}
\label{table:Photos}
\vspace{-3mm}
\end{table}
%208
\begin{table}
\centering
\begin{tabular}{| c |}
\hline
\includegraphics[width=0.4\textwidth]{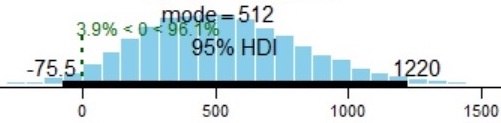} \\
Photos Category\\
\hline
\end{tabular}
\captionof{figure}{Regions of Practical Equivalence for the difference of means ($\mu_{1}-\mu_{2}$) for our participants and the Android marketplace in selection of photo apps (RQ3)}
\label{fig:photos}
\vspace{-5mm}
\end{table}
\vspace{-3mm}
\subsection{Discussion}
Can we use the most common indicator of privacy on the Internet-- a lock--  to communicate aggregate information about mobile app privacy risk? If so, would this change the choices made by participants in the marketplace? There is no a priori answer. Information about privacy and security risks could be ignored or unwelcome. Studies in risk communication have shown that individuals find risk more acceptable if the exposure to the risk is voluntary, and when the individual exposed is capable of avoiding the risk or freely choosing it~\cite{garg2012end}. That is, shifting the nexus of control may actually increase aggregate risk-taking; the perception of control increases data sharing~\cite{stutzman2013silent}. In privacy, this response is called the `control dilemma'~\cite{brandimarte2013misplaced}. To address these questions in the context of mobile apps, we asked the research questions described above.

First (RQ1), can we confirm that our group of participants are indistinguishable from the Google's PlayStore users as a whole when presented with apps that had the same kind of functionality and the same privacy ratings? We used the Flashlight category for this purpose. We found that the selection of apps under this condition was indistinguishable from a random sample of app selections in the PlayStore. Our next question (RQ2) is if the participants would make different choices compared to that of Google's PlayStore users in the presence of variable ratings given the same functionality. For this question, we used the Weather category. The results from this category showed that choices in our experiment are significantly different from those made in the Google PlayStore, thus offering a high level of confidence that the ratings influenced user decision-making. Finally, we ask (in RQ3) if we can be confident that the participants' decisions are different from Google's PlayStore users when the functionality and the privacy ratings both vary. We used Games and Photos as the categories to answer this question. The Photos category indicates that the inclusion of privacy ratings, even with marginal rating differences, results in a different distribution of apps selected. In Games, we have a low level of confidence that the participants' decisions are different and can not conclude the parameters are changed.

In summary, we built a functional app store with real, accurately rated apps and added visual indicators for aggregate risk using the padlock icon. The functional app store simulation made it possible for us to compare the choices of the participants directly with those of people using Google's PlayStore. It is true that using a functional app store with real-world applications meant that we were not able to adequately control for biases, including ordering, familiarity, and reputation. However, we are confident that our participants' choices when there was no variance in privacy is indistinguishable from those of the PlayStore at large, providing confidence in the representativeness of our sample that is difficult to obtain in a traditional controlled laboratory experiment. 
\vspace{-4mm}
\section{Acknowledgement}
This research was supported in part by the National Science Foundation under CNS 1565375, Cisco Research Support, and the Comcast Innovation Fund. Any opinions, findings, and conclusions or recommendations expressed in this material are those of the author(s). They do not necessarily reflect the views of the U.S. Government, NSF, Cisco, Comcast, Indiana U, or the University of Denver.
\vspace{-4mm}
\begin{small}

\section{References}
\begingroup
\renewcommand{\section}[2]{} 
\bibliographystyle{abbrv}
\bibliography{Android}
\endgroup
\end{small}
\end{document}